# Molecular Dynamics Simulations of Ballistic Penetration of Pentagraphene Sheets

David L. Azevedo[1,2,3], Rafael A. Bizao[3], and Douglas S. Galvao[3]

[1]Instituto de Física, Universidade de Brasília, Brasília-DF, 70910900, Brazil.

[2]UnB Planaltina College, Universidade de Brasília, CEP 73345-10, Planaltina-DF, Brazil

[3]Applied Physics Department and Center of Computational Engineering and Science, University of Campinas - UNICAMP, Campinas-SP 13083-959, Brazil.

## ABSTRACT

*The superior mechanical properties and low density of carbon nanostructures make them promising ballistic protection materials, stimulating investigations on their high-strain-rate behavior. Recent experiments and simulations revealed graphene possesses exceptional energy absorption properties. In this work, we analyzed through fully atomistic molecular dynamics simulations the ballistic performance of a carbon-based material recently proposed named penta-graphene. Our results show that the fracture pattern is more spherical (no petals formation like observed for graphene). The estimated penetration energy for pentagraphene structures considered here was of 37.69 MJ/Kg, far superior to graphene (29.8 MJ/Kg) under same conditions. These preliminary results are suggestive that pentagraphene could be an excellent material for ballistic applications.*

## INTRODUCTION

With the discovery of carbon nanotubes and graphene, a large number of theoretical and experimental studies have been carried out in part due to their remarkable mechanical properties [1,2]. Although most of these studies are for planar and cylindrical structures, it is also possible to obtain other structure topologies, such as coils (helicoidal type) and carbon nanocones (CNCs). Recently, Zhang *et al.* [3] proposed a new class of carbon-based nanostructures, named 'penta-graphene'. Pentagraphene membranes are quasi-bidimensional structures formed by densely compacted pentagons (instead of hexagons in graphene) carbon atoms in $sp^3$ and $sp^2$ hybridized states – see Figures 1 and 2).

Recently [1,2,4,5], many works addressed the ballistic properties of graphene membranes. Pentagraphene share some interesting mechanical properties [3,6] with graphene. A natural question is about the similarities and differences of pentagraphene membranes under ballistic impact in comparison to graphene ones. One of the objectives of the present work is to address these questions.

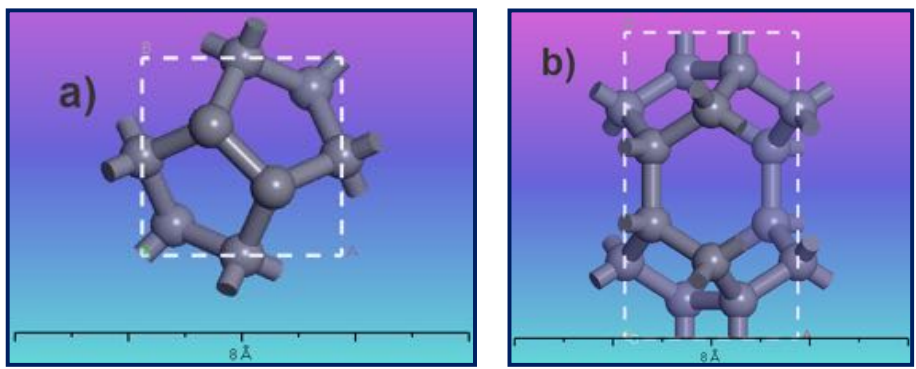

Figure 1: Unit cell of a monolayer pentagraphene sheet seen from (a) XY and; (b) XZ planes.

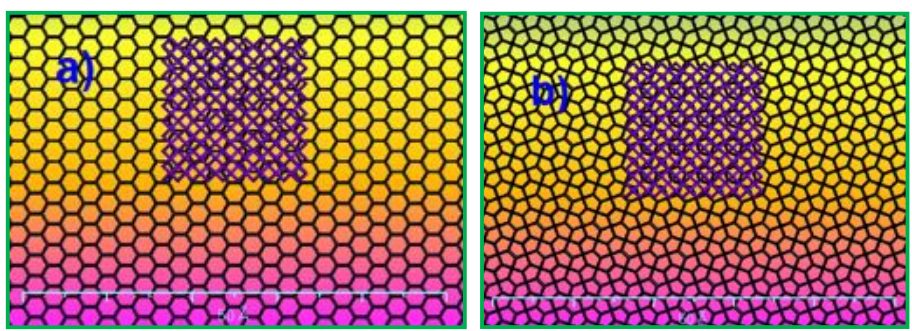

Figure 2: Replicated unit cell of (a) graphene, and; (b) pentagraphene. For reference a diamond fragment (in blue) is on each structure.

## METHODS

We carried out fully atomistic molecular dynamics (MD) simulations using the reactive force field ReaxFF [7]. ReaxFF is paremeterized directly from ab initio calculations and constrasted agaist experimental values. For hydrocarbon systems, the differences between the predicted heat of formation values for theory and experiments are ~3.0 Kcal/mol [9].

We simulated impact dynamics of diamond particle composed of 68,372 carbon atoms packed into r ~ 4.5 nm radius sphere with an initial kinetic energy of $2.45 \times 10^6$ Kcal/mol, and a pentagraphene membrane formed of 177,504 carbon atoms totalizing a system of 245,876 atoms (Figure 2). Initially, the membrane was thermalized in a NPT ensemble keeping null pressure along x and y directions. Then, we performed a second thermalization in a NVT ensemble at 300 K for the whole system, and finally using a NVE ensemble the diamond ball was shot against the membrane with an initial speed of 5 Km/s. The used time step integration was 0.02 fs with ReaxFF parameters of Muller *et al.* [9]. In the Figure 2, we present a MD snapshot of the initial system configuration.

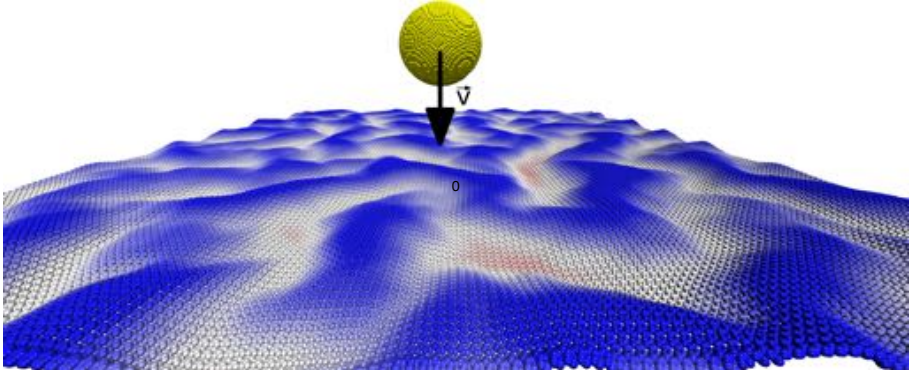

Figure 2: MD snapshot of the initial system configuration. The membrane is colored in relation to the z-coordinate, from red (below zero) to blue (above). The observed rugosity is due to thermal fluctuations at 300K.

## DISCUSSION

In Figure 3, we present the results for the impact particle velocity and the corresponding kinetic energy absorption during the first 4 (four) picoseconds of the MD simulation. As we can see from the Figure, after impact there is a significant velocity decrease, but even fracturing the membrane the kinetic energy absorption was only 14%.

In Figure 4 we present MD snapshots immediately after particle perforation. As we can see from this Figure the produced hole is approximately spherical. In contrast to graphene [1,2,4,5] no petal-like structures are formed, validating the model that the petal formation was a consequence of the hexagonal lattice. In the case of pentagraphene, the significant buckling and the pentagonal topology prevents the petal formation.

The specific perforation energy can be estimated using the following equation:

$$d_1 = \frac{\Delta E}{\rho A_p t} \ (1)$$

In which $\Delta E$, $A_p$, $\rho$, t are respectively the kinetic energy dissipated after perforation, projectile cross section area projected on the sheet, density, and thickness of the single layer, which is equivalent to the ratio of kinetic energy variation over the sheet mass with the same cross section area of the incident particle. Substituting the following parameters for the present case in equation (1): $r = 45$ Angstroms (nano particle radius), mass of sheet for the projected area $\rho A_p t = 6.151 \times 10^{-23}$ Kg, and $\Delta E = 2.318 \times 10^{-15}$ Joules (kinect energy lost after perforation), we obtain $d_{1penta} = 37.69$ MJ/Kg for pentagraphene, which is larger than the corresponding value for graphene ( $d_{1grafeno} = 29.8$ MJ/Kg) for graphene under same conditions [4,5], which suggest that pentagraphe could be a promising material for ballistic applications.

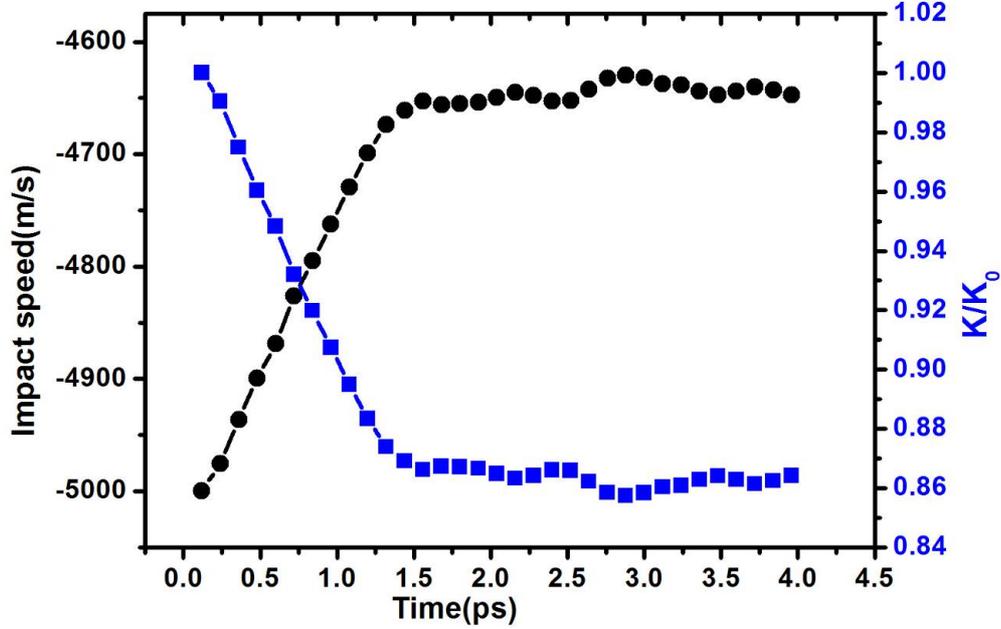

Figure 3: Black circles represent impact velocity of the nanoparticle as a function of the simulation time. Blue squares represent the corresponding kinetic energy absorption. For the systems considered here, the maximum absorbed kinect energy was ~ 14%.

**CONCLUSION**

We have investigated the mechanical properties and fracture patterns of armchair and zigzag-like pentagraphene nanotubes (PGNTs). Pentagraphene is a quasi-2D (buckled) carbon allotrope composed of pentagons (and not hexagons, like in graphene). Our results show that PGNT fracture patterns is chirality dependent. This is due to the different alignments with the directions of the applied unitension (axial) tension (see Figures 2 and 3), which results in armchair tubes breaking at earlier stages than the corresponding zigzag-like ones. The estimated PGNT Young's modulus values for the structures considered here were on average 800 GPa, which is 36% lower than conventional carbon nanotubes.

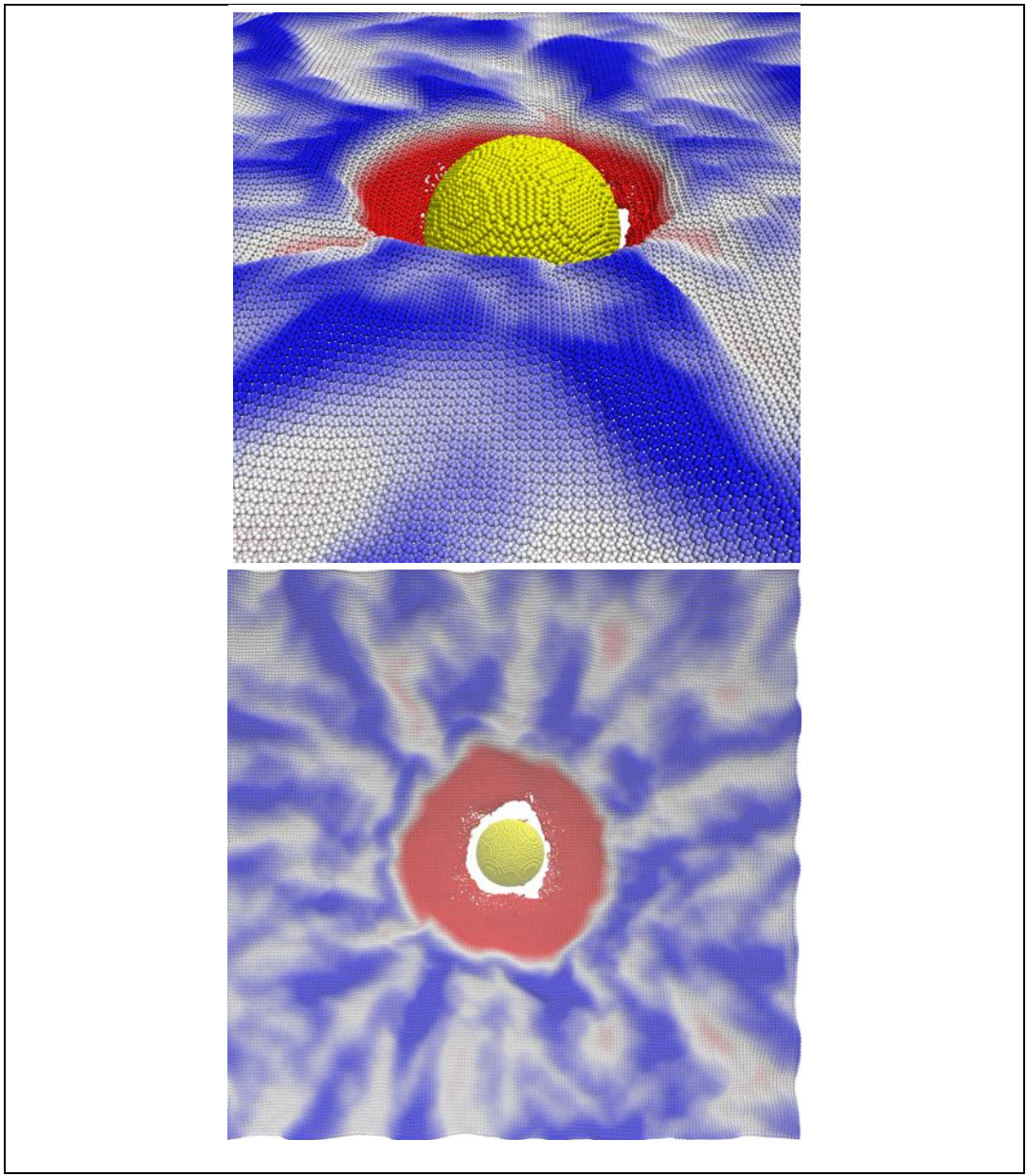

Figure 4: *(a) Stress strain curve for PGNT with chirality zigzag-like (9,0) and armchair-like (9,9). For the PGNT zigzag-like the critical strain is 20.5% and the PGNT armchair-like is 22.6%. (b)* Young's Modulus values for zigzag-like and armchair-like PGNTs. (c) Critical strain $\sigma_c$ values for zigzag-like and armchair-like PGNTs.

## ACKNOWLEDGEMENTS

This work was supported in part by the Brazilian agencies CAPES, CNPq, FAPESP, UnB/DPI. DSG acknowledges the Center for Computational Engineering and Sciences at Unicamp for financial support through the FAPESP/CEPID Grant #2013/08293-7.